\documentclass{vgtc}                          

\graphicspath{{Figures/}{./}}                 

\usepackage{times}                     
\usepackage{mathptmx}                  
\usepackage{enumitem}                  

\let\origcite\cite
\renewcommand{\cite}[1]{\mbox{\origcite{#1}}}

\onlineid{0}

\vgtccategory{Research}

\vgtcinsertpkg

\title{A Data-Centric Perspective on Tree Visualizations}

\renewcommand{\thefootnote}{\fnsymbol{footnote}}
\author{Manling Yang\thanks{e-mail: \{manling.yang, alexandra.scott, chris.ahn, daniel.jakab, suyang.li, mingwei.li, remco.chang\}@tufts.edu.} %
\and Alexandra Scott$^{*}$\thanks{e-mail: amscott@mitre.org.} %
\and Chris Ahn$^{*}$ %
\and Daniel Jakab$^{*}$ %
\and Suyang Li$^{*}$ %
\and Mingwei Li$^{*}$ %
\and Remco Chang$^{*}$}
\affiliation{%
  $^{*}$\scriptsize Tufts University\quad
  $^{\dagger}$\scriptsize The MITRE Corporation}

\abstract{
Tree visualization (TreeVis) techniques span diverse designs.
Existing taxonomies organize them by visual characteristics such as layout dimensionality, edge representation, and node alignment.
However, this visual-centric perspective can obscure structural similarities and make it difficult to determine whether differences arise from data structures or visual encodings.
We investigate TreeVis techniques from a data-centric perspective grounded in Prepared Tables, the final data state prior to visual encoding.
Using TreeVis.net, we curate 133 two-dimensional techniques and characterize each by the object records and attribute roles required before encoding.
Our analysis shows that the corpus is more concentrated at the prepared-data level than a visual reading would suggest.
The techniques collapse to a small set of recurring object combinations and schemas.
Many techniques across TreeVis representation categories share the same schema, suggesting that much of the apparent diversity of TreeVis designs lies in visual representation rather than fundamentally different pre-encoding data requirements.
Prepared Table schemas therefore support reasoning about structural equivalence, sufficiency, and difference across TreeVis designs.
}

\begin{document}

\maketitle
\renewcommand{\thefootnote}{\arabic{footnote}}

\section{Introduction}
TreeVis have developed into a substantial design space for representing hierarchical data, including node-link diagrams, treemaps \cite{shneiderman1992tree}, icicle plots \cite{kruskal1983icicle}, containment-based designs, and hybrids such as elastic hierarchies \cite{zhao2005elastic}.
This diversity has been documented through resources such as TreeVis.net, a web-based survey of TreeVis techniques organized by dimensionality, edge representation, and node alignment \cite{schulz2011treevis}.
Its continued growth, now reaching 341 techniques, underscores the need for comparison methods that characterize the design space beyond visual appearance.

Existing references and taxonomies effectively describe how techniques differ visually, for example through explicit versus implicit edge representation, containment, subdivision, or hybrid composition \cite{schulz2011treevis,scheibel2020taxonomy}.
However, visual categories do not directly explain whether two techniques require different pre-encoding data structures or instead realize the same structure through different visual forms.
This limits our ability to assess whether a technique introduces a substantive structural contribution or primarily re-expresses an existing structure through a new visual encoding.
For example, a node-link tree and a treemap-like representation may visualize the same hierarchy but require different pre-encoding structures. A node-link representation can make parent-child relations explicit as edge records, such as \texttt{Node(id)} and \texttt{Edge(id, source, target)}. In contrast, a treemap-like representation can encode hierarchy through node records with parent references, such as \texttt{Node(id, parent\_id, value)}, where \texttt{value} is required when area represents quantity. Both begin from hierarchical data, but the objects and attributes explicit before visual encoding can differ.

To study tree visualizations at the data level rather than the appearance level, we draw on the Information Visualization Reference Model \cite{card1999readings} and Wu and Chang's notion of the Prepared Table, the final state of the data table prior to visual encoding \cite{wu2024design}.
We use Prepared Table schemas to describe the pre-encoding requirements imposed by TreeVis techniques, including the object types and attributes required for a representation to be constructed and interpreted.
This perspective enables comparison in terms of structural equivalence, sufficiency, and difference at the data-centric level.

Using TreeVis.net as a corpus, we curate 133 two-dimensional TreeVis techniques and encode each technique as a Prepared Table schema.
We analyze the corpus through three schema-level relations: equivalence, which identifies techniques that share object combinations or exact schemas; sufficiency, which identifies when one schema already satisfies another; and difference, which identifies techniques that introduce additional object types or specialized semantic attributes.
The analysis shows that visual diversity and pre-encoding structural diversity are not aligned one-to-one: many visually distinct techniques share the same schema, reuse relations extend beyond exact equivalence, and structural departures are comparatively limited and patterned.
These findings position Prepared Table schemas as a useful lens for separating visual variation from data-level variation in TreeVis design.

Overall, the contributions of this work are as follows:
\begin{itemize}[itemsep=0.15em, topsep=0.25em, parsep=0pt]
    \item We introduce Prepared Table schemas to compare TreeVis techniques by their required data objects and attributes.
    \item We curate 133 two-dimensional TreeVis techniques from TreeVis.net and encode them as Prepared Table schemas for data-centric comparison.
    \item We show that TreeVis techniques exhibit strong schema-level concentration, substantial sufficiency-based reuse, and limited patterned structural differences despite visual variation.
\end{itemize}
\section{Related Work}
\label{sec:related_work}

Prior work organizes the tree visualization design space primarily by visual form.
TreeVis.net \cite{schulz2011treevis}, Li et al.'s feature-based analysis \cite{li2015exploring}, GoTreeScape \cite{li2022gotreescape}, and treemap taxonomies \cite{scheibel2020taxonomy} characterize techniques by dimensions such as representation, alignment, visual features, grammar, or treemap variants.
These works provide valuable visual-centric accounts, but they do not directly compare the pre-encoding objects and attributes required by different techniques.
Our work complements them by comparing TreeVis techniques through their Prepared Table schemas.

Visualization models show that design depends not only on visual encoding, but also on how data is transformed before encoding. The Information Visualization Reference Model separates data transformation, visual mapping, and view transformation \cite{card1999readings}, while Data State Reference Model formalizes visualization techniques as transformations across representational states \cite{chi2000taxonomy}. Wu and Chang further define the Prepared Table as the final data table prior to encoding \cite{wu2024design} (see \Cref{fig:model}). Related systems such as APT \cite{mackinlay1986automating}, Show Me \cite{mackinlay2007show}, Draco \cite{moritz2018formalizing}, VizML \cite{hu2019vizml}, and Voyager \cite{wongsuphasawat2015voyager} formalize visualization design over structured data, but mainly for generation or recommendation rather than comparison of techniques.

Building on this work, we compare TreeVis techniques through Prepared Table schemas to analyze structural equivalence, sufficiency, and difference.

\section{Prepared Table}
\label{sec:prepared_table}
This section defines Prepared Tables as the final data state before visual encoding (\Cref{ssec:prepared_table_definition}) and specifies their schemas by data objects and attributes (\Cref{ssec:prepared_table_schema}).

\subsection{Definition}
\label{ssec:prepared_table_definition}
\begin{figure}
    \centering
    \includegraphics[width=0.99\columnwidth]{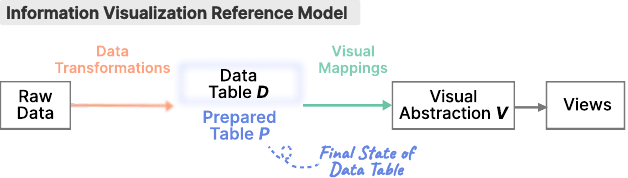}
    \vspace{-0.2cm}
    \caption{
        The Prepared Table as the final data state before visual encoding, capturing the schema-level requirements of a visualization technique within the Information Visualization Reference Model \cite{card1999readings}.
    }
    \label{fig:model}
    \vspace{-0.2cm}
\end{figure}

The Information Visualization Reference Model distinguishes data transformations from visual mappings, making clear that visual encoding operates over a transformed data state rather than directly over raw data \cite{card1999readings,chi2000taxonomy,munzner2009nested}.
Following Wu and Chang, we refer to this final pre-encoding data state as a Prepared Table $P$ \cite{wu2024design}.
The Prepared Table therefore serves as the structural interface between upstream data transformation and downstream visual encoding (see \Cref{fig:model}).
It determines which data objects and attributes have been made explicit before encoding, and therefore what can be directly mapped to visual marks and channels for encoding \cite{bertin1983semiology,mackinlay1986automating,munzner2014visualization}.

This distinction is useful for comparing TreeVis because techniques that appear visually different may still require the same pre-encoding structure.
For example, two techniques may use different layouts or visual encodings while relying on the same set of node records and attributes before encoding.
Conversely, two visually related techniques may require different Prepared Tables if one introduces additional derived records or specialized attributes before encoding.
Thus, our focus is not only on what a tree visualization looks like, but on what data objects and roles must already exist before visual encoding begins.

Because published TreeVis papers rarely expose the data table used by an implementation, we do not attempt to reconstruct implementation-specific tables.
Instead, we use the Prepared Table as a conservative abstraction of pre-encoding requirements, inferred from the representation and its accompanying description.
The next subsection makes this abstraction operational by defining Prepared Table schemas in terms of required objects and attributes.

\subsection{Prepared Table Schema}
\label{ssec:prepared_table_schema}
\begin{figure*}
    \centering
    \includegraphics[width=0.95\textwidth]{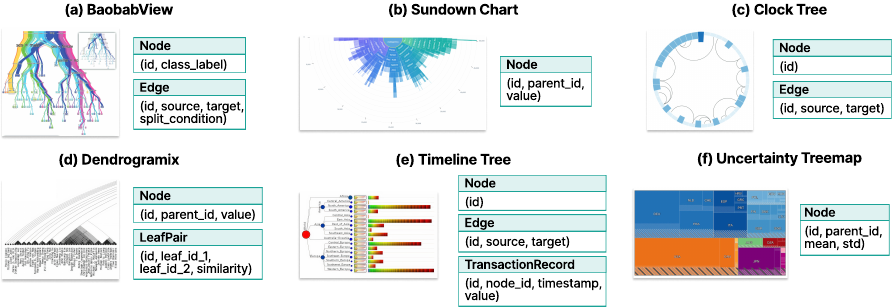}
    \vspace{-0.2cm}
    \caption{Examples of Prepared Table schemas for TreeVis techniques. Panels (a)-(c) show representative Explicit, Implicit, and Hybrid techniques: \textit{BaobabView} \cite{van2011baobabview}, \textit{Sundown Chart} \cite{woodburn2019interactive}, and \textit{Clock Tree} \cite{li2020gotree}. Panels (d)-(f) illustrate structural differences introduced by additional object types or specialized attributes, including \textit{LeafPair} records in \textit{Dendrogramix} \cite{blanch2015dendrogramix}, transaction records in \textit{Timeline Tree} \cite{burch2008timeline}, and uncertainty attributes in \textit{Uncertainty Treemap} \cite{sondag2020uncertainty}.}
    \label{fig:structural_patterns}
    \vspace{-0.2cm}
\end{figure*}

We use the term \emph{Prepared Table schema} to refer to a data-type-level description of the objects and attributes required by a visualization design.
The schema is defined at the object-and-attribute level rather than at the instance level \cite{codd1970relational,elmasri2016fundamentals}.
It therefore does not describe how many nodes, edges, or other records appear in a particular visualization.
Instead, it specifies what kinds of objects and attributes a technique assumes before visual encoding.
We record two schema elements for each technique:
\begin{itemize}[leftmargin=*]
    \item \textbf{Objects:} analysis objects instantiated as first-class records in the Prepared Table, such as nodes, edges, or derived records summarizing node relationships.
    \item \textbf{Attributes:} the semantic information attached to objects that is required for the representation to be constructed and interpreted.
\end{itemize}
Together, objects and attributes define the Prepared Table schema of a technique.
For corpus coding, we apply an object-identification rule to make schema modeling less arbitrary. Tree topology is represented as \texttt{parent\_id} on \texttt{Node} when parent-child relations only define containment, subdivision, or membership among nodes; \texttt{Edge} is instantiated separately when links are first-class encoded relations or carry their own semantic attributes. Additional object types are introduced only when the represented records cannot be attached unambiguously to a single node or edge, such as pairwise leaf relations \cite{blanch2015dendrogramix} or transaction records \cite{burch2008timeline}.
We distinguish encoded attributes from derived attributes using a necessity test: an attribute is included only when its removal would prevent the representation from being constructed or interpreted according to the data semantics documented in the source material. Attributes are excluded when computable from already included objects and attributes during layout or rendering. For example, in \textit{Sundown Chart} (\Cref{fig:structural_patterns} (b)), \texttt{value} is retained because it encodes quantitative hierarchy, while angular positions, radii, and arc boundaries are excluded as layout results. In \textit{BaobabView} (\Cref{fig:structural_patterns} (a)), \texttt{split\_condition} is retained because it encodes decision-tree semantics on edges rather than edge layout or routing.
Positions of visual marks, control points, and intermediate layout variables are excluded when they result from layout computation rather than representing pre-encoding data requirements \cite{battista1998graph,munzner2014visualization}.
For attribute naming, we do not treat all node or edge labels as a single undifferentiated field, because doing so would collapse different semantic requirements into the same schema attribute. Generic quantitative magnitudes are normalized as \texttt{value}, while technique-specific semantic attributes, such as uncertainty information, are retained with descriptive names when necessary and supported by the source material \cite{gortler2017bubble,sondag2020uncertainty}.

This definition makes the Prepared Table schema a comparative abstraction rather than an implementation trace. A specific implementation of a visualization technique may introduce additional information for layout, interaction, or computational efficiency, which we omit because it is not strictly necessary for rendering the final visualization. By limiting the schema to include only the required objects and attributes, we compare tree visualizations based on their pre-encoding data requirements rather than on system-specific implementation details.

\section{Analysis}
\label{sec:analysis}
This section describes how we curated and encoded the TreeVis corpus as Prepared Table schemas (\Cref{ssec:corpus_encoding}), and reports the findings from our analysis (\Cref{ssec:structural_patterns}).

\subsection{Corpus and Encoding}
\label{ssec:corpus_encoding}
Following prior corpus-construction practice in tree visualization research \cite{li2015exploring}, we examine techniques in TreeVis.net published after the early 1980s. We restrict the corpus using the following criteria:

\begin{itemize}[leftmargin=*]
    \item \textbf{Two-Dimensional Representation.}
    Three-dimensional and immersive representations introduce spatial navigation and viewpoint-dependent factors beyond the schema-level focus of this analysis.
    \item \textbf{Representation-Level Distinctness.}
    We retain only techniques that introduce distinct representation-level designs rather than variants in layout algorithms and interaction designs.
    \item \textbf{Sufficient Documentation.}
    We retain only techniques whose pre-encoding structural requirements can be reliably inferred from documentation.
\end{itemize}

Applying these criteria to the 341 techniques in TreeVis.net yielded a final corpus of 133 techniques. The two-dimensional filter removed 67 three-dimensional techniques and 4 techniques with hybrid dimensions, leaving 270. We then excluded 128 techniques that did not introduce a distinct representation-level design, including techniques that primarily contributed layout algorithms (e.g., \cite{sondag2018stable, harel2000algorithm}) or interaction techniques (e.g., \cite{blanch2007browsing, tominski2006fisheye}) while preserving an existing visual representation. Finally, we excluded 9 techniques whose documentation did not provide enough evidence to infer the required objects and attributes without speculation (e.g., \cite{Eichhorn2006, Sukla2005}).

Each retained technique was encoded as a Prepared Table schema by recording the objects and attributes required before visual encoding, using the definitions in \Cref{ssec:prepared_table_schema} as the shared coding basis. To improve consistency, three authors followed a shared labeling protocol, reviewed representative examples against the source documentation, and then divided the retained techniques among the three coders. Seven retained techniques were marked as uncertain during coding and resolved collectively by checking the associated papers, with TreeVis.net entries and other design descriptions used as supplementary evidence when available. All seven uncertain cases were resolved and retained in the final corpus. The first author subsequently reviewed the final labels for corpus-wide consistency.

\subsection{Structural Patterns}
\label{ssec:structural_patterns}
We compare the Prepared Table schemas through three schema-level relations: structural equivalence, structural sufficiency, and structural difference.
Here, $T$ denotes a technique's Prepared Table schema; $T_1$ and $T_2$ denote the schemas of compared techniques:

\begin{itemize}[leftmargin=*]
    \item \raisebox{-0.15em}{\includegraphics[height=1em]{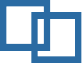}} \textbf{Structural Equivalence.} We consider two types of equivalence, one at the object level, and the other at the schema level:
    \begin{itemize}[leftmargin=1.5em, label=\scalebox{0.8}{$\circ$}, topsep=0pt]
    \item \textbf{Object-Level Equivalence}: Two techniques that use the same objects are considered to be structurally equivalent at the object-level. For example, BaobabView (\Cref{fig:structural_patterns} (a)) and ClockTree (\Cref{fig:structural_patterns} (c)) are said to have object equivalence because they both include Node and Edge objects.
    \item \textbf{Schema-Level Equivalence}: Two techniques are considered equivalent at the schema level when they share the exact schema, including both objects and all attributes.
    \end{itemize}

    \item \raisebox{-0.15em}{\includegraphics[height=1em]{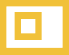}} \textbf{Structural Sufficiency.}
    A Prepared Table schema $T_2$ is structurally sufficient for a technique with schema $T_1$ when every object and attribute required by $T_1$ is also present in $T_2$.
    Sufficiency indicates that data prepared for $T_2$ can support a technique with schema $T_1$ without additional restructuring.

    \item \raisebox{-0.15em}{\includegraphics[height=1em]{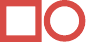}} \textbf{Structural Difference.}
    Techniques differ when their Prepared Table schemas require different objects or specialized attributes that cannot be reduced to the same prepared structure.
    Difference indicates that moving between techniques would require substantive restructuring before encoding, such as introducing derived records or specialized attributes.
\end{itemize}

These comparisons provide an analytical lens for understanding the design space of tree visualizations.
Equivalence asks when visually distinct techniques rely on the same pre-encoding structure.
Sufficiency asks when one prepared structure can support the requirements of another.
Difference asks when a technique introduces a genuinely different structural requirement before encoding.
Together, these comparisons allow us to separate visual diversity from pre-encoding data-structure diversity and to examine whether apparent design variation corresponds to different Prepared Table schemas or to different visual representations of the same schema.

\smallskip
\noindent
\raisebox{-0.15em}{\includegraphics[height=1em]{Figures/equivalence.pdf}} \textbf{Structural equivalence accounts for much of the observed corpus structure.}
This pattern clarifies when visual differences do not imply different prepared data requirements. If visually distinct techniques share object types or exact schemas, their differences can be understood primarily as visual, layout, or encoding choices rather than changes in the data objects and attributes required before encoding. In this sense, equivalence helps identify schema-level reuse across the TreeVis design space.
We examine equivalence at two levels: object-level equivalence provides a coarse view of which record types a technique requires, while schema-level equivalence further tests whether those shared objects also carry the same semantic attribute roles.

\textit{At the Object-Level Equivalence, the corpus collapses to six groups.}
To characterize structural equivalence at the object-level, we count the distinct combinations of object types required by each technique, independent of attributes associated with those objects.
The resulting distribution is concentrated: the corpus contains only six object-type combinations across 133 techniques.
Most techniques require only \texttt{Node}: 88 techniques (66.2\%), while 41 techniques (30.8\%) require both \texttt{Node} and \texttt{Edge}.
Together, these two cases account for 129 techniques (97.0\% of the corpus), indicating that object-type variation explains a small portion of the design space.
The remaining four combinations each occur once and introduce one object type beyond \texttt{Node} and \texttt{Edge}: \texttt{Instance}, \texttt{LeafPair} (e.g., \Cref{fig:structural_patterns} (d)), \texttt{Silhouette}, or \texttt{TransactionRecord}.

\begin{figure}[t]
    \centering
    \includegraphics[width=\columnwidth]{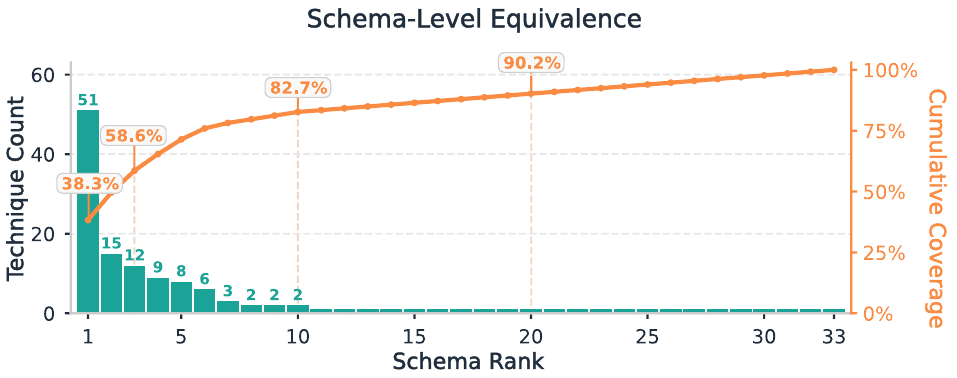}
    \vspace{-0.5cm}
    \caption{%
        Distribution of the 33 exact schemas ranked by frequency (bars, left axis) and their cumulative corpus coverage (curve, right axis).
        The top schema alone, \texttt{Node\{id, parent\_id, value\}}, accounts for 51 techniques (38.3\%); the top three and top ten schemas reach 58.6\% and 82.7\%, respectively.
    }
    \label{fig:schema_tail}
    \vspace{-0.2cm}
\end{figure}

\textit{At the Schema-Level Equivalence, the corpus collapses to 33 exact schemas, but most techniques are covered by a small number of schemas.}
We count distinct schemas shared across tree visualization techniques and compute the cumulative coverage of the most frequent schemas (\Cref{fig:schema_tail}), where cumulative coverage is defined as the percentage of techniques represented by a given set of schemas.
We find 33 distinct exact schemas, an increase from the six object-type combinations. However, the distribution is highly skewed: a few schemas account for most techniques, while many schemas occur only once.
The most frequent schema, \texttt{Node\{id, parent\_id, value\}}, covers 38.3\% of the corpus; the top three schemas cover 58.6\%, the top five cover 71.4\%, and the top ten cover 82.7\%.
As shown in \Cref{fig:schema_tail}, coverage gains diminish rapidly after the first few ranks: the remaining 23 schemas each appear in exactly one technique, forming a pronounced long tail.
Schema concentration also appears within individual TreeVis representation categories, each with a distinct dominant schema.
In the Implicit category, \texttt{Node\{id, parent\_id, value\}} alone covers 57.5\% of techniques (50/87); in the Explicit category, \texttt{Edge\{id, source, target\}; Node\{id, value\}} accounts for 25.0\% (9/36); and in the Hybrid category, \texttt{Node\{id, parent\_id, type, value\}} accounts for 20.0\% (2/10).

Together, these results show that TreeVis representation categories and Prepared Table schemas are not in a one-to-one correspondence: visually distinct techniques often share the same schema, which explains why only 33 schemas suffice to describe this large and visually diverse corpus.

\smallskip
\noindent
\raisebox{-0.15em}{\includegraphics[height=1em]{Figures/containment.pdf}}
\textbf{Structural sufficiency reveals reuse relations beyond exact equivalence.}
Exact equivalence identifies techniques that share the same Prepared Table schema, whereas sufficiency identifies schemas that can satisfy the requirements of other techniques without additional restructuring.
We define the \emph{sufficiency radius} of a schema $T_2$ as the number of techniques with schema $T_1$ whose required objects and attributes are already present in $T_2$.
Because \texttt{Node}-only and \texttt{Node+Edge} schemas require fundamentally different object types, sufficiency relations are confined within each structural family: no single schema bridges both.
Within the \texttt{Node}-only family, the schema \texttt{Node\{id, parent\_id, type, value\}} achieves the largest sufficiency radius, covering 77 techniques (57.9\% of the corpus) across the Implicit and Hybrid categories.
This means that a dataset prepared to this schema already satisfies the pre-encoding requirements of 77 techniques without any further data restructuring, even though those techniques may differ substantially in their visual representations.

\smallskip
\noindent
\raisebox{-0.15em}{\includegraphics[height=1em]{Figures/difference.pdf}} \textbf{Structural difference is limited in frequency and form.}
To identify techniques that impose different pre-encoding requirements, we identify schemas that introduce object types or specialized semantic roles beyond the dominant schema patterns.
Only 15 techniques (11.3\%) depart from the dominant schemas, clustering into a small set of recurrent mechanisms.
Four techniques introduce an object type beyond \texttt{Node} and \texttt{Edge}, such as \textit{Dendrogramix} \cite{blanch2015dendrogramix}, which materializes \texttt{LeafPair} records (see \Cref{fig:structural_patterns} (d)).
Six techniques add specialized attributes to \texttt{Edge} beyond connectivity, such as \textit{BaobabView} \cite{van2011baobabview}, which uses \texttt{split\_condition} (see \Cref{fig:structural_patterns} (a)).
Two techniques introduce explicit multi-tree coordination through attributes such as \texttt{tree\_id} or \texttt{correspondence\_id}, including \textit{Barcode Tree} \cite{li2019barcodetree} and \textit{Google+Ripples} \cite{viegas2013google}.
Three techniques introduce temporal semantics through time-related attributes:
\textit{Timeline Tree} \cite{burch2008timeline} uses \texttt{timestamp} and
\texttt{transaction\_id} (see \Cref{fig:structural_patterns}(e));
\textit{Ripple Presentation} \cite{ishihara2006ripple} encodes \texttt{time}
on edges; and \textit{DiskTree} \cite{chi1998visualizing} records temporal state.
Two techniques, including \textit{Uncertainty Treemap} \cite{sondag2020uncertainty}, incorporate statistical attributes such as \texttt{mean} and \texttt{std} (see \Cref{fig:structural_patterns} (f)).
These patterns show that structural difference in the corpus is not arbitrary, but organized around recurring schema-level changes.

\section{Discussion}
\label{sec:discussion}
\noindent

Our use of Prepared Table schemas shifts comparison from how tree visualization techniques look to what must be structurally available before visual encoding. Many visually distinct techniques, including techniques from different TreeVis representation categories, rely on the same small set of objects and attributes. This suggests that much of the TreeVis design space varies through visual design, encoding, or layout choices rather than through fundamentally different pre-encoding data requirements.

A smaller group of techniques introduces new pre-encoding commitments, such as uncertainty attributes, multi-tree correspondence identifiers, or transaction records. These techniques support capabilities tied to specific data, contexts, or tasks because they require additional information to be explicit before encoding. However, relatively few techniques fall into this category, suggesting that most TreeVis designs remain general-purpose rather than committing to specialized data semantics. For example, when uncertainty is not modeled as a distinct semantic attribute, it is often treated as a generic value attribute and encoded through standard channels such as node size or edge width.
Prepared Table schemas complement visual taxonomies by making structural commonalities and differences explicit. They provide a basis for comparing tree visualization techniques at the level of their underlying data requirements rather than representational form alone.

\smallskip
\noindent
\textbf{Limitations and Future Work.}
Our analysis is incomplete: we do not analyze attribute measurement types, such as categorical or quantitative attributes, as a corpus-level dimension.
Prepared Table schemas abstract away visual encoding, layout, interaction, and perceptual factors, so structural equivalence does not imply functional or evaluative equivalence. Techniques with the same schema may differ in readability, scalability, task fit, or user interpretation, while structural difference only indicates a change in what must be explicit before encoding. Future work should combine schema-level analysis with visual and empirical evaluation to examine when these structural relations matter for design and use.

\section{Conclusion}
\label{sec:conclusion}
This paper examined TreeVis techniques through Prepared Table schemas: the object types and attribute roles required before visual encoding.
Using 133 two-dimensional techniques from TreeVis.net, we showed that the design space is far more concentrated at the Prepared Table schema level than a visual reading would suggest.
Many visually different techniques, including those from different TreeVis representation categories, share the same schema, while others depart through a limited set of recurrent structural mechanisms.
Together, these results show that visual diversity and pre-encoding structural diversity do not align one-to-one.
The Prepared Table schema perspective complements existing visual taxonomies by supporting reasoning about structural equivalence, sufficiency, and difference before encoding.
\\
\\
\noindent\textbf{Supplementary Materials.}\\
The corpus and coding materials are available in the \href{https://osf.io/crq7w/overview?view_only=bb27a187f7b24fcaa683af79e6b0fddc}{OSF repository}.

\acknowledgments{
The authors thank the anonymous reviewers and paper chairs for their constructive feedback.
This work was supported in part by the National Science Foundation under Grant No. OAC-2118201.
}

\bibliographystyle{abbrv-doi-hyperref}

\bibliography{references}
\end{document}